\documentclass[10pt,twocolumn]{article}

\usepackage[letterpaper,margin=1in,columnsep=0.24in]{geometry}
\usepackage[T1]{fontenc}
\usepackage[utf8]{inputenc}
\usepackage{mathptmx}
\usepackage{amsmath,amssymb}
\usepackage{booktabs}
\usepackage{array}
\usepackage{tabularx}
\usepackage{graphicx}
\usepackage{url}
\usepackage{balance}
\usepackage[numbers,sort&compress]{natbib}
\usepackage[hidelinks]{hyperref}
\usepackage{titlesec}
\usepackage[singlelinecheck=false,justification=centering]{caption}
\graphicspath{{figures/}}

\renewcommand{\thesection}{\Roman{section}}
\renewcommand{\thesubsection}{\Alph{subsection}}
\titleformat{\section}{\normalfont\large\bfseries}{\thesection.}{0.5em}{}
\titleformat{\subsection}{\normalfont\normalsize\bfseries}{\thesubsection.}{0.5em}{}
\titlespacing*{\section}{0pt}{1.0ex plus 0.5ex}{0.6ex}
\titlespacing*{\subsection}{0pt}{0.8ex plus 0.4ex}{0.4ex}
\title{\vspace{-0.35in}\textbf{A Multi-Analyst LLM Pipeline for Auditable Rule Discovery Across 68 Public Physiological Corpora}}
\author{D\=ovy Paukstys\\Komori Care, LLC\\Waterford, Virginia, USA\\dovy@komoricare.com}
\date{7/7/26}

\begin{document}

\twocolumn[
\maketitle
\begin{abstract}
Open physiological corpora are heterogeneous: they use different sensors, labels, sampling rates, recording settings, and clinical endpoints. They can support detector design, but they do not directly specify which detector rules should be built for a new contactless monitoring platform. We report a controlled four-analyst large-language-model (LLM) workflow for converting 68 public physiological corpora, screened for commercial-use compatibility, into an auditable library of candidate rule shapes for prospective validation. Four independent commercial LLM families read the corpus documentation under a controlled prompt and produced 695 candidate rule markers (top-markers). Deduplication retained 649 rule records; a threshold-bounds audit then flagged 51 sanity violations for clamping or curator review. Cross-corpus consolidation produced 436 unique rule shapes. Gate-tagging against two hard invariants, native target-hardware channel availability and no multi-night per-patient personalization, identified 94 build-now detector components across four detector-family buckets. The pipeline does not produce a validated clinical detector. It produces an auditable engineering cascade in which analyst disagreement, threshold checks, curator review, and automated continuous-integration (CI) checks route literature-derived rules toward prospective hardware validation.
\end{abstract}
\begin{center}
\end{center}
\vspace{0.08in}
]

\begin{figure*}[t]
\centering
\includegraphics[width=0.88\textwidth]{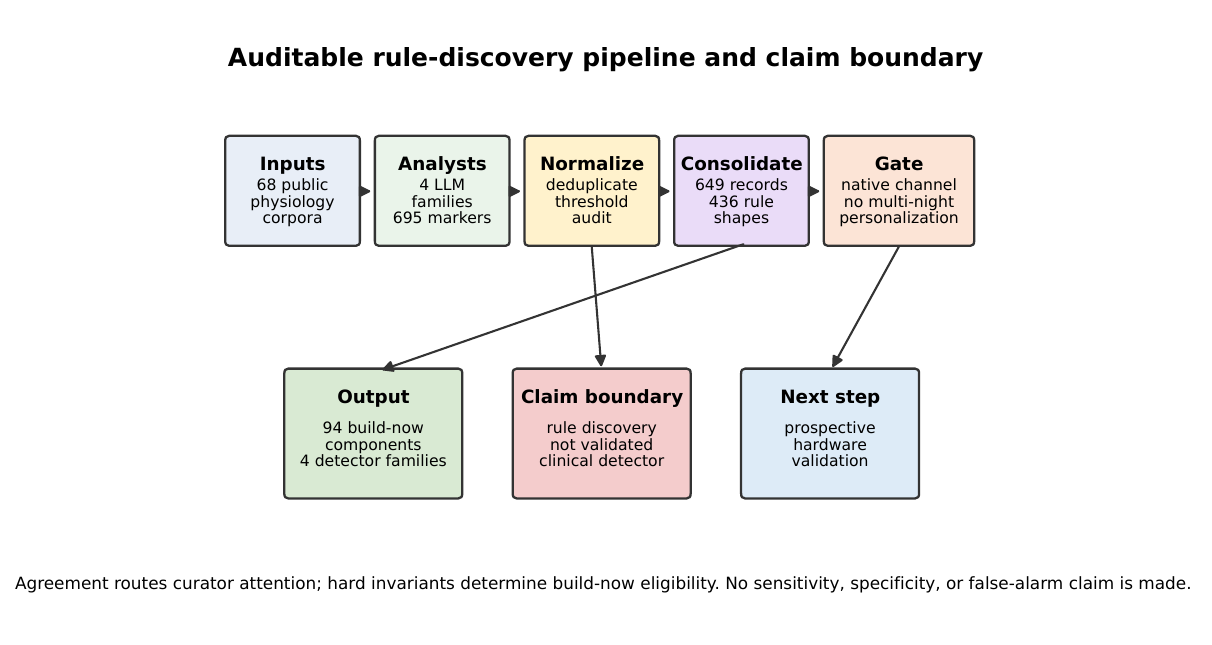}
\caption{The workflow converts heterogeneous public-corpus documentation into gated detector-component candidates. The output is a rule-discovery artifact, not a validated clinical detector.}
\label{fig:pipeline-boundary}
\end{figure*}

\section{Introduction}
\label{sec:introduction}

Single-corpus detector work is brittle: a rule that performs well on one dataset may fail after a sensor change, a label change, or a deployment-context change. We needed a detector-rule library for a contactless nocturnal-monitoring platform that uses millimeter-wave radar, WiFi channel-state information (CSI), low-resolution thermal sensing, on-device acoustic features, and ambient sensors (light, temperature, humidity, barometric pressure, and CO\textsubscript{2}). Public corpora cover many relevant physiological phenomena, but to our knowledge none of the public corpora we screened were recorded on this target hardware, and none directly specify an engineering rule library. The exact per-modality acquisition parameters (radar center frequency and chirp/frame structure, thermal resolution and frame rate, CSI sampling, on-chip detection settings, and host-visible outputs) are part of the proprietary hardware configuration summarized only at the channel-availability level in Table~\ref{tab:disclosure-boundary}; this paper evaluates whether a rule maps to a channel that exists on the platform, not the channel's calibrated performance. The deployment target is nocturnal in-bed monitoring, but the source corpora span both nocturnal and non-nocturnal recordings (e.g., daytime cardiac and postural datasets), which is one reason hardware gates are applied per rule.

The literature base is broad but fragmented. The corpus and prior-art pool includes SeizeIT2 and its validation work \cite{seizeit2,seizeit2val}, CHB-MIT \cite{chbmit}, SUDEP and nocturnal/wearable seizure-detection literature \cite{mortemus,vilella,beniczky2013,beniczky2018,onorati,ilae,arends,vanandel,bruno}, cardiac and public physiological repositories \cite{mitbih,physionet}, neonatal EEG seizure data \cite{helsinki}, radar vital-sign work \cite{radarbins}, clinical large-language-model (LLM) prior work \cite{medpalm}, posture-response physiology \cite{postural1,postural2}, and WiFi CSI vital-sign work \cite{wificsi}. These sources are valuable, but they mix clinical labels, sensor modalities, patient populations, and feasibility assumptions. Table~\ref{tab:related-positioning} positions this work against these adjacent sources of evidence, stating what each contributes and what it does not provide.

This paper makes three claims. First, no single public corpus can scope a multi-sensor contactless detector library. Second, a single analyst, human or model, risks systematic blind spots when extracting rules across dozens of heterogeneous corpora. Third, disagreement between independent analysts can be used as a triage signal: it exposes contested extractions and routes curator attention, but it does not itself establish correctness.

Our contribution is not a new clinical detector and not a benchmark of LLM vendors. The contribution is a constrained engineering cascade: four independent analyst outputs are normalized, deduplicated, audited for threshold sanity, gate-tagged against target hardware, checked by continuous-integration (CI) invariants, and routed toward prospective validation. The observed cascade was:

\begin{center}
\small
695 analyst top-markers $\rightarrow$ 649 retained rule records $\rightarrow$ 436 unique rule shapes $\rightarrow$ 94 build-now detector components.
\end{center}

Fig.~\ref{fig:pipeline-boundary} summarizes the pipeline and claim boundary. The contributions are:
\begin{itemize}
\item A controlled multi-analyst extraction workflow for turning heterogeneous physiological-corpus documentation into normalized candidate rule shapes.
\item A staged audit cascade---deduplication, threshold-bounds checking, curator review, and CI-enforced hardware invariants---that prevents literature-derived rules from becoming unvalidated product claims.
\item A disagreement analysis showing that analyst agreement routes attention but does not determine implementability.
\item A disclosure pattern for reporting aggregate provenance while withholding proprietary rule definitions, thresholds, prompts, and gate rationales.
\end{itemize}

\begin{table*}[t]
\centering
\caption{Positioning Against Adjacent Sources of Evidence}
\label{tab:related-positioning}
\footnotesize
\renewcommand{\arraystretch}{1.12}
\begin{tabularx}{0.96\textwidth}{p{0.18\textwidth}p{0.23\textwidth}p{0.23\textwidth}X}
\toprule
Evidence line & What it contributes & What it does not provide & Role in this paper \\
\midrule
Public physiological corpora \cite{chbmit,mitbih,physionet,helsinki} & Heterogeneous labels, sensors, sampling rates, and physiological phenomena & A target-hardware detector library or deployable contactless performance & Input evidence for candidate rule-shape extraction, not direct validation. \\
Wearable/nocturnal seizure detection studies \cite{beniczky2013,beniczky2018,onorati,ilae,arends,vanandel,bruno,seizeit2,seizeit2val} & Clinically relevant event families, endpoints, and failure modes & Transfer guarantees to radar/CSI/thermal/audio contactless sensing & Source of rule families and cautionary constraints, routed through hardware gates. \\
Radar and WiFi sensing work \cite{radarbins,wificsi} & Sensor-specific physiological feasibility under defined hardware assumptions & Proof that extracted rules work on the target hardware or operating point & Supplies modality priors; build-now status still requires native-channel and no-personalization gates. \\
Clinical LLM work \cite{medpalm} & Evidence that LLMs can encode and manipulate biomedical knowledge & A rule-discovery audit cascade or detector engineering workflow & Motivates careful use of LLM analysts while keeping them non-authoritative. \\
This paper & An auditable extraction, consolidation, and gate-tagging method over 68 corpora & Sensitivity, specificity, false-alarm rate, latency, or clinical utility & Produces candidate components for prospective hardware validation. \\
\bottomrule
\end{tabularx}
\end{table*}

We deliberately do not report sensitivity, specificity, false-alarm rate, or latency. Those metrics require prospective hardware validation against medical-grade ground truth.

\section{Method}
\label{sec:method}

\subsection{Corpus Selection and Disclosure Boundary}
\label{subsec:corpus}

The pipeline ingests 68 public physiological corpora whose use was screened for commercial compatibility. Each corpus required: (i) a public release or controlled public archive with documented access terms; (ii) documented sensor modality and sampling rate; and (iii) at least one nocturnal-monitoring-relevant label or phenomenon, such as heart rate, respiration, body position, sleep stage, seizure event, autonomic response, bed-exit behavior, or postictal (after-seizure) recovery.

The 68 corpora span the modality and label families cited above (\cite{chbmit,mitbih,physionet,helsinki,seizeit2,seizeit2val,radarbins,wificsi} are representative public anchors); the cited datasets are representative anchors, and the full enumerated registry (corpus key, license note, and modality tag per entry) is provided in the reviewer bundle as a redacted manifest rather than inline, because the complete list and prioritization expose product scope.

The full corpus registry, license notes, and derived rule artifacts are version-controlled. The registry can support audit of the input set, but public disclosure is limited because raw rule definitions, thresholds, prompt text, provider-specific disagreements, and gate rationales expose proprietary product logic. This manuscript therefore reports aggregate counts, methods, audit stages, and disclosure boundaries rather than releasing the production rule library.

\begin{table*}[t]
\centering
\caption{Disclosure Boundary for Auditability Without Publishing Proprietary Rule Logic}
\label{tab:disclosure-boundary}
\footnotesize
\renewcommand{\arraystretch}{1.12}
\begin{tabularx}{0.96\textwidth}{p{0.18\textwidth}X X}
\toprule
Layer & Public or reviewer-safe disclosure & Withheld from public release \\
\midrule
Input set & Corpus count, corpus-selection criteria, license-screen method, modality/label categories & Full internal registry if it exposes product prioritization or unreleased licensing notes \\
Analyst runs & Aggregate marker counts, analyst-family independence design, prompt-control principles, run hashes or redacted audit card & Raw prompts, provider-specific raw outputs, provider-specific disagreements, per-pair analyst-comparison records (battle output), scorecards \\
Rule records & Aggregate transitions: 695 $\rightarrow$ 649 $\rightarrow$ 436 $\rightarrow$ 94; category-level agreement table & Raw rule definitions, thresholds, rule text, detection logic, curator rationales, gate-decision logs \\
Quality gates & Threshold-bounds audit count, native-channel invariant, no-personalization invariant, CI invariant concept & Exact threshold-bounds vocabulary, product-specific hardware gate logic, proprietary CI rules \\
Validation path & Statement that all components require prospective target-hardware validation & Any implication of validated clinical performance before prospective data exist \\
\bottomrule
\end{tabularx}
\end{table*}

This paper reports no primary human-subjects data. All analyses were performed on previously published, de-identified public corpora or public corpus documentation governed by their original release terms.

\subsection{Four-Analyst Design}
\label{subsec:four-analyst}

Four independent commercial LLM families served as analyst pipelines. We do not rank vendors and do not report provider-specific performance. Independence at the model-family level was the design requirement. Each analyst used the same controlled prompt, which specified a normalized rule representation, citation requirements, and a build-readiness gate taxonomy. Pairwise analyst-comparison records (internally, the \emph{battle output}) capture which family surfaced which marker and are withheld from public release.

The analysts were used as parallel extractors, not authorities. Agreement prioritizes review, but it does not determine implementability. A 4/4-agreement rule can be gated out if it violates the native-sensor invariant. A 1/4-agreement rule can remain eligible if it passes invariants and curator review. The workflow uses disagreement to route scrutiny rather than treating LLM agreement as truth.

\subsection{Normalization, Deduplication, and Threshold Audit}
\label{subsec:normalization}

The four analysts produced 695 top-markers (candidate rules each analyst surfaced as most salient for a corpus). Deduplication retained 649 rule records (93.4\%) using signal/modality canonicalization, time-window normalization, citation provenance, and rule-shape matching. A separate threshold-bounds audit over the retained records flagged 51 sanity violations for clamping or curator review. The threshold audit is part of the contribution: it shows where model-derived or literature-derived thresholds are unsafe to carry forward without review.

Cross-corpus consolidation then produced 436 unique rule shapes. The deduplication and correction code is unit-tested and runs in CI when outputs are regenerated.

For example, four differently worded seizure-literature rules describing post-burst stillness can normalize to one canonical motion-modality rule shape: short post-burst stillness, thresholded against a session-local baseline inside the event window. The threshold is drawn from the controlled vocabulary or routed through the threshold-bounds audit, not accepted from one analyst's wording. The session-local baseline is not a multi-night personalized baseline.

\subsection{Gate-Tagging and Hard Invariants}
\label{subsec:gate-tagging}

Each of the 436 unique rule shapes was tagged by build-readiness: build-now component, tier-2 probe, feature prior, clinician review, or archive.
The final gate-tagging classifier (internal pipeline milestone Phase 0.3) assigned the distribution in Table~\ref{tab:gate-distribution}:

\begin{table}[t]
\centering
\caption{Final gate-tagging distribution (internal pipeline milestone Phase 0.3).}
\label{tab:gate-distribution}
\footnotesize
\begin{tabular}{lr}
\toprule
Action & Rule shapes \\
\midrule
Build-now component & 94 \\
Tier-2 probe & 97 \\
Feature prior & 81 \\
Clinician review & 100 \\
Archive & 64 \\
\midrule
Total & 436 \\
\bottomrule
\end{tabular}
\end{table}

The non-build-now actions are: \emph{tier-2 probe} (plausible but requires a follow-up feasibility probe before promotion), \emph{feature prior} (informs a feature but is not itself a detector rule), \emph{clinician review} (clinically consequential, needs domain sign-off), and \emph{archive} (incompatible with the target hardware or out of scope).

Two invariants are asserted before a rule can remain in the build-now bucket:

\begin{enumerate}
\item Native sensor channel: every build-now component must read from a sensor channel that exists natively on the target hardware. No proxy substitution.
\item No per-patient personalization: no build-now component may require multi-night per-patient baseline learning or individualized user-profile setup.
\end{enumerate}

These are engineering gates, not clinical validation. A pull request or regeneration step that leaves a build-now rule violating either invariant fails. Personalization-heavy or proxy-heavy literature is retained as a probe, feature prior, or future regulatory claim rather than promoted to a build-now component.

\subsection{Curator Review and Failure Modes}
\label{subsec:curator}

Each of the 436 unique rule shapes was reviewed by one human curator against a fixed rubric: semantic distinctness, literature support, native-channel availability, threshold sanity, and absence of multi-night personalization. The curator decision is an engineering triage decision, not clinical ground truth.

Observed failure modes included hallucinated citations, threshold drift, over-claimed build-readiness, label-overlap confusion, model-version drift, and prompt sensitivity. Mitigations include citation verification, threshold-bounds audit, CI invariant checks, artifact hashes for run provenance, and curator review before final gate-tagging.

A secondary AI spot check of 50 retained rule-map records found that about half required probe-first or gate-clarification treatment, mainly because of native-channel, contactless-proxy, or baseline-scope risk. This supports the conservative gate taxonomy but is not an independent inter-rater study (two or more independent human reviewers scoring the same items).

Table~\ref{tab:audit-ladder} lays out the audit ladder: for each question the pipeline answers, it pairs the artifact or metric, the audit check, and the claim boundary that bounds it.

\begin{table*}[t]
\centering
\caption{Audit Ladder and Claim Boundary}
\label{tab:audit-ladder}
\footnotesize
\renewcommand{\arraystretch}{1.12}
\begin{tabularx}{0.96\textwidth}{p{0.19\textwidth}p{0.22\textwidth}p{0.25\textwidth}X}
\toprule
Question & Artifact or metric & Audit check & Claim boundary \\
\midrule
Can heterogeneous corpora be converted into candidate rule shapes? & 68 screened corpora, 695 analyst top-markers, 649 retained records & Corpus screen, normalized representation, deduplication trace & Shows extraction coverage and curation flow, not correctness of every candidate rule. \\
Does analyst agreement determine survival? & Agreement table and inclusion-rate comparison across 4/4, 3/4, 2/4, and 1/4 markers & Chi-square comparison and category-level disagreement review & Agreement is a triage signal; it is not treated as truth or implementability. \\
Are unsafe thresholds routed out of the build path? & Threshold-bounds audit with 51 flagged sanity violations & Clamp or curator-review route before consolidation and gating & Demonstrates a safety filter; does not validate any threshold on target hardware. \\
Can target-hardware constraints be enforced? & 436 rule shapes gate-tagged into five action buckets; 94 build-now components & Native-channel and no-multi-night-personalization invariants checked in CI & Build-now means engineering-ready for prospective testing, not clinically validated. \\
Is the pipeline auditable under proprietary constraints? & Hashes, run metadata, stage counts, redacted one-corpus audit-card pattern & Disclosure boundary in Table~\ref{tab:disclosure-boundary} & Supports provenance audit without releasing product rule logic. \\
\bottomrule
\end{tabularx}
\end{table*}

\section{Results}
\label{sec:results}

\subsection{Cascade}
\label{subsec:cascade}

\begin{figure*}[t]
\centering
\includegraphics[width=0.98\textwidth]{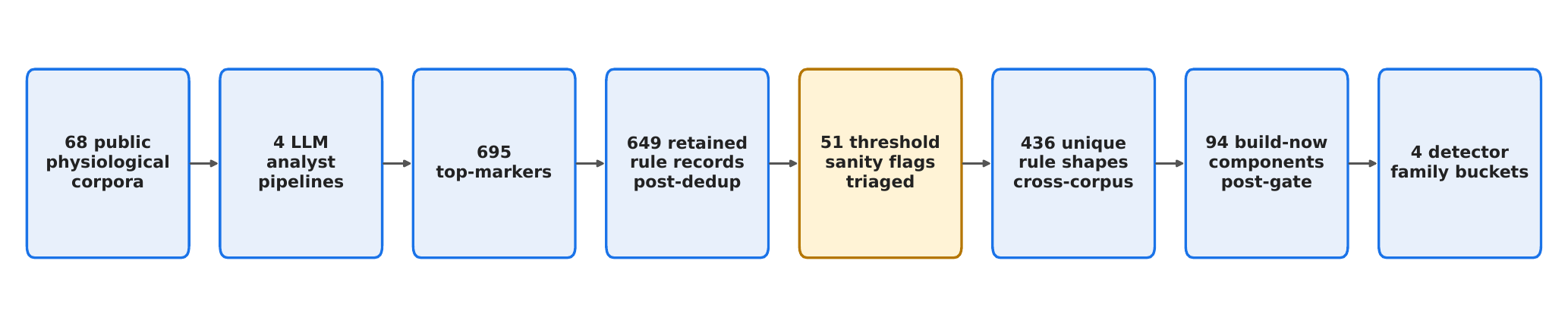}
\caption{Four-analyst rule-discovery cascade. Counts shown are aggregate engineering artifacts, not detector-performance metrics.}
\label{fig:cascade}
\end{figure*}

The end-to-end cascade is summarized in Fig.~\ref{fig:cascade} and Table~\ref{tab:cascade}.

\begin{table}[t]
\centering
\caption{End-to-end cascade counts.}
\label{tab:cascade}
\footnotesize
\begin{tabularx}{\columnwidth}{@{}p{0.30\columnwidth}rX@{}}
\toprule
Stage & Count & Interpretation \\
\midrule
Analyst top-markers & 695 & Candidate markers surfaced by analyst pipelines \\
Retained rule records & 649 & Records retained after deduplication, with threshold defects routed for correction/review \\
Threshold-bounds flags & 51 & Sanity violations flagged for clamping or curator review (a subset of the 649, not a separate reduction step) \\
Unique rule shapes & 436 & Consolidated cross-corpus rule shapes \\
Build-now components & 94 & Native-channel, no-personalization components after the build-readiness gates \\
\bottomrule
\end{tabularx}
\end{table}

Reduction is curation, not loss. Non-build-now rules remain useful as tier-2 probes, feature priors, clinician-review candidates, or archived incompatible rules.

\subsection{Build-Now Detector-Family Buckets}
\label{subsec:families}

The 94 build-now components map to four detector-family buckets (Table~\ref{tab:families}) that share an event interface and temporal-state model:

\begin{table}[t]
\centering
\caption{Build-now components by detector-family bucket.}
\label{tab:families}
\footnotesize
\begin{tabular}{lr}
\toprule
Detector-family bucket & Components \\
\midrule
Autonomic surge with failed recovery & 62 \\
Postictal respiratory compromise & 24 \\
Bed-exit after high-movement event & 6 \\
Postictal recovery risk & 2 \\
\midrule
Total & 94 \\
\bottomrule
\end{tabular}
\end{table}

Autonomic surge with failed recovery (a sharp heart-rate/respiration rise that does not return to baseline) contains the largest native-channel component pool. Postictal respiratory compromise is retained as a separate family because it depends on respiratory-channel readiness and has a distinct failure mode. Bed-exit after high-movement event is small but well-defined (a single, directly measurable bed-exit event). Postictal recovery risk is clinically important but currently has only two build-now components; it should not be treated as an independently validated detector family before prospective validation.

The shared temporal model is baseline $\rightarrow$ motor burst $\rightarrow$ rhythmic clonic (repetitive jerking) or high-movement interval $\rightarrow$ post-event stillness/recovery. The same event interface lets conservative build-now components coexist with probe-first literature-derived candidates without turning unvalidated probes into product claims.

\subsection{Inter-Analyst Disagreement}
\label{subsec:disagreement}

A rule marker is a candidate rule surfaced by one or more analyst pipelines. Agreement level is the number of analyst pipelines that surfaced the marker.

\begin{table}[t]
\centering
\caption{Inter-analyst agreement across 695 top-markers. Library-inclusion rate did not differ across agreement levels (chi-square=0.52, df=3, p=0.91).}
\label{tab:agreement}
\footnotesize
\begin{tabular}{lrr}
\toprule
Agreement & Markers & Fraction \\
\midrule
4/4 & 256 & 36.8\% \\
3/4 & 211 & 30.4\% \\
2/4 & 110 & 15.8\% \\
1/4 & 118 & 17.0\% \\
\midrule
Total & 695 & 100.0\% \\
\bottomrule
\end{tabular}
\end{table}

Two-thirds of top-markers (67.2\%) carried 3/4 or 4/4 agreement. Single-AI-only markers were 17.0\%. Per-agreement-level retention rate (retained category-assignments / top-markers at that level) was essentially uniform: 93.0\%, 93.4\%, 93.6\%, and 94.9\% for 4/4, 3/4, 2/4, and 1/4 respectively (chi-square = 0.52, p = 0.91, df = 3; inclusion computed over the 650 marker-category assignments spanning the 649 retained records, since one record carries two category labels). Agreement did not predict survival. Instead, agreement changed where curator attention was directed.

\begin{table}[t]
\centering
\caption{Agreement by marker category, over the 650 category assignments on the 649 retained records (cf. Table~\ref{tab:agreement}, which is over all 695 top-markers). Seizure vs Cardiac single-AI rate differs (Fisher exact OR=2.97, p=0.0012).}
\label{tab:category-agreement}
\scriptsize
\setlength{\tabcolsep}{2.1pt}
\resizebox{\columnwidth}{!}{%
\begin{tabular}{lrrrrrr}
\toprule
Category & 4/4 & 3/4 & 2/4 & 1/4 & Total & \% single-AI \\
\midrule
Cardiac & 133 & 107 & 45 & 40 & 325 & 12.3\% \\
Screening & 6 & 6 & 8 & 3 & 23 & 13.0\% \\
Thermal & 6 & 4 & 2 & 2 & 14 & 14.3\% \\
Autonomic & 11 & 17 & 4 & 6 & 38 & 15.8\% \\
Audio & 2 & 1 & 2 & 1 & 6 & 16.7\% \\
Respiratory & 28 & 20 & 20 & 14 & 82 & 17.1\% \\
Movement & 16 & 15 & 9 & 11 & 51 & 21.6\% \\
Sleep & 12 & 10 & 5 & 8 & 35 & 22.9\% \\
Seizure & 24 & 16 & 8 & 20 & 68 & 29.4\% \\
Composite & 0 & 1 & 0 & 7 & 8 & 87.5\% \\
\bottomrule
\end{tabular}}
\end{table}

The agreement-by-category breakdown is given in Table~\ref{tab:category-agreement}. Unlike Table~\ref{tab:agreement}, which is computed over all 695 top-markers, this category breakdown is computed over the 649 retained records; the per-row \%~single-AI is therefore on the retained basis and is not directly comparable to the 17.0\% single-AI figure reported for the full 695-marker set. Total mapped marker-category assignments were 650 across 649 retained records because one record was assigned to two categories. Cardiac markers dominated the cohort and showed the highest consensus. Seizure markers showed more disagreement, consistent with transfer difficulty across patients, sensors, and event definitions. The Cardiac-versus-Seizure difference was statistically significant: seizure markers were about three times more likely than cardiac markers to be single-AI-only (2x2: Seizure 20 single-AI vs 48 multi-AI of 68; Cardiac 40 single-AI vs 285 multi-AI of 325; Fisher's exact test, odds ratio OR = 2.97, p = 0.00124, where the odds ratio is the relative odds of a marker being single-AI-only). Audio and Composite categories are small-n descriptive categories and should not be overinterpreted.

Agreement triages. Invariants determine implementability.

\subsection{Audit Trail}
\label{subsec:audit}

Every cascade transition is attributable to a recorded decision: corpus selection, analyst output, deduplication, threshold audit, gate-tagging, CI invariant check, and curator review. Table~\ref{tab:audit} lists the artifact logged at each stage and its reviewer value.

\begin{table*}[t]
\centering
\caption{Audit artifacts logged at each stage.}
\label{tab:audit}
\footnotesize
\begin{tabularx}{0.96\textwidth}{p{0.20\textwidth}X X}
\toprule
Stage & Artifact logged & Reviewer value \\
\midrule
Corpus selection & Corpus manifest hash, license note, modality tags & Fixes the input set and commercial-use screen \\
Analyst run & Prompt identifier, run date, model-family/version string when exposed, output hash & Bounds model and prompt provenance \\
Normalization & Vocabulary version, deduplication code version, threshold-bounds report & Makes the 695 $\rightarrow$ 649 $\rightarrow$ 436 transition inspectable \\
Gate-tagging & Classifier version, gate-decision log, CI invariant report & Makes build-now vs probe-first decisions inspectable \\
Curator review & Rubric version, decision timestamp, rationale category & Exposes human judgment instead of hiding it \\
\bottomrule
\end{tabularx}
\end{table*}

As detailed in the disclosure boundary (Table~\ref{tab:disclosure-boundary}), the rule library, raw prompts, gate-decision logs, and pipeline code are withheld. To demonstrate audit structure without publishing rule substance, the accompanying ancillary/reviewer bundle includes a redacted one-corpus audit card, audit-artifact manifest, public corpus-registry schema, claim-to-evidence matrix, reproducibility protocol, and release-boundary note. These files disclose corpus key, run metadata, artifact hash, aggregate marker counts, stage-level artifacts, and redaction policy while withholding marker names, thresholds, detection methods, provider-specific positions, curator rationales, and gate logic. They support auditability and reviewer inspection, not public bit-for-bit reproduction of the rule library.

\section{Discussion}
\label{sec:discussion}

This paper describes a methodology, not a detector. The 94 build-now components are scoping-stage rule shapes and component candidates, not alert-ready clinical detectors. Performance metrics will be reported only after prospective hardware validation against medical-grade ground truth. The fact that a rule is literature-supported, multi-analyst-surfaced, and invariant-clean is not evidence of sensitivity, specificity, false-alarm rate, or clinical utility.

The main result is that disagreement was useful, but not in the naive way. Agreement did not predict inclusion into the retained library. Instead, disagreement exposed where curator review mattered most. Cardiac rules were relatively canonical. Seizure and Composite markers were more contested. That is precisely the setting in which a single-reader extraction would be easiest to overtrust.

The novelty is also narrower and more defensible than ``LLMs read biomedical papers.'' LLMs have already been studied for clinical reasoning \cite{medpalm}. The contribution here is a constrained signal-processing workflow: independent LLM analysts, normalized rule-shape representation, threshold-bounds auditing, hard target-hardware invariants, and gate-tagged routing to prospective validation.

The gate taxonomy prevents literature findings from becoming unvalidated product claims. For example, bedside radar work can report strong vital-sign performance under specific capture geometry, frame-rate, windowing, and range-bin assumptions \cite{radarbins}. In this workflow such results become probe-first or hardware-readiness-gated until the same assumptions are tested on target hardware. Similarly, wearable or multimodal seizure-detector studies \cite{beniczky2013,beniczky2018,onorati,ilae,arends,vanandel,bruno} inform rule families without implying that their performance transfers to contactless sensing.

The paper has important limitations. First, the final gate-tagging review was performed by a single human curator. That is acceptable for engineering triage but not for clinical ground truth; an independent human domain-review spot check would strengthen the work. Second, the raw rule library and prompts are withheld (see Table~\ref{tab:disclosure-boundary}), so the paper claims auditable provenance rather than public bit-for-bit reproducibility. Third, prompt sensitivity and commercial model-version drift remain risks. Fourth, the source corpora differ in label quality, sensor placement, recording context, and patient population. Fifth, four analysts were useful here, but the study does not prove that four is the optimal number of analyst pipelines in every domain.

These limitations define the next validation steps. The candidate components must be tested prospectively on target hardware, against independent ground truth, with pre-specified operating points and false-alarm accounting. The audit framework should also be tested with an independent human reviewer or second-curator sample, and the controlled-disclosure package should expose enough artifact hashes and redacted audit cards for reviewers to verify that the cascade exists without releasing proprietary rule definitions.

\section{Conclusion}
\label{sec:conclusion}

We built a four-analyst LLM pipeline over 68 public physiological corpora screened for commercial-use compatibility. The pipeline produced 695 analyst top-markers, retained 649 rule records, consolidated them into 436 unique rule shapes, and identified 94 build-now detector components after native-channel and no-personalization invariants. The output is not a validated clinical detector. It is an auditable, conservative rule-discovery cascade for moving heterogeneous open-corpus evidence toward prospective contactless-hardware validation.

\section*{Data, Code, and Artifact Availability}

The article source package includes an ancillary/reviewer bundle under \path{anc/paper_A_reviewer_release_bundle_2026-07-07/}. The bundle contains a redacted one-corpus audit card, audit-artifact manifest, public corpus-registry schema, claim-to-evidence matrix, controlled reproducibility protocol, and release-boundary note. The production rule library, raw prompts, full analyst outputs, thresholds, gate rationales, and implementation code are proprietary and are not publicly released. Confidential review of selected hashes or redacted artifacts can be supported without disclosing rule substance.

\section*{Acknowledgement}

No vendor sponsored this work. The four analyst pipelines are independent commercial LLM families; no per-vendor comparison is made and the methodology is intended to be vendor-agnostic.

D. Paukstys is founder of Komori Care, LLC. The production rule library and implementation artifacts are proprietary (disclosure boundary in Table~\ref{tab:disclosure-boundary}). This paper reports the rule-discovery methodology and aggregate audit cascade, not a validated detector or released product.

Manuscript drafting and editorial review were assisted by AI tools (OpenAI ChatGPT and Anthropic Claude) for organization, wording, critique, and polishing across the abstract, introduction, methods, discussion, availability, and limitation sections. This editorial assistance is distinct from the four-analyst LLM pipeline described in the research methodology. The author is solely responsible for all data, analyses, citations, methods, claims, and conclusions.

\balance

\end{document}